\documentclass[preprint,12pt]{elsarticle}

\usepackage{amsmath}
\usepackage{amssymb}
\usepackage[titletoc,toc,title]{appendix}
\usepackage{graphicx}
\usepackage{float}
\usepackage{bm}
\usepackage{xcolor}
\usepackage{booktabs}
\usepackage{hyperref}
\usepackage{diagbox}
\usepackage{braket}
\usepackage{nicefrac}
\usepackage{dsfont}
\usepackage{comment}

\usepackage{listings}
\usepackage{xcolor} 

\definecolor{river}{HTML}{517C96}

\journal{Computer Physics Communications}

\newcommand{\1}{\hat{\mathbb{1}}}
\newcommand{\Iop}{\hat{I}}
\newcommand{\Bop}{\hat{\mathcal{B}}}
\newcommand{\rhoop}{\hat{\rho}}
\newcommand{\rhom}{\rhoop_\text{m}}
\newcommand{\rhoth}{\rhoop_\text{th}}

\newcounter{bla}

\journal{Computer Physics Communications}

\begin{document}

\begin{frontmatter}

\title{Computational framework for quantum state tomography of spin ensembles}

\author[a]{Jiwoo Seo}
\author[a,b]{Vesna F. Mitrovi\'{c} \corref{cor}}

\cortext[cor]{Corresponding author {e-mail address:}  vemi@brown.edu}
\address[a]{Department of Physics, Brown University, Providence, RI 02912, USA}
\address[b]{Brown Center for Theoretical Physics and Innovation (BCTPI), Brown University, Providence, RI 02912, USA}

\begin{abstract}
We present Tomography-NMR, an open-source Python package that reconstructs quantum density matrices from spectroscopic measurement data.
The package implements a complete analysis pipeline for two-qubit quantum state tomography based on the product operator formalism: raw time-domain signals are Fourier-transformed into frequency-domain spectra, spectral peak intensities are mapped to expansion coefficients of the density matrix, and the full quantum state is reconstructed.
Three integration methods are provided for different use cases: direct peak height measurement and fixed-parameter numerical integration require no theoretical reference and are suited to unknown states, achieving fidelities of approximately 98\% on a benchmark set of known states, while a systematic parameter optimization against a known target state achieves reconstruction fidelities exceeding 99\% for the same benchmark states.
While the detailed theoretical framework for quantum state tomography is well established, the practical procedures for extracting density matrices from experimental spectra are often inadequately documented in the literature and obscured within proprietary software.
This package addresses that gap by providing a fully transparent, reproducible implementation of every analysis step, from spectral preprocessing to density matrix visualization.
The software has been validated on experimentally prepared two-qubit states measured via nuclear magnetic resonance (NMR) spectroscopy of coupled $^{31}$P nuclei.
Average reconstruction fidelities range from 0.975 to 0.995 across a benchmark set of 20 two-qubit states, including the computational basis states, Bell states, and the outputs of three fundamental quantum gates (CNOT, H, and T).
Although developed for NMR, the modular architecture facilitates adaptation to other spectroscopic platforms and alternative measurement protocols.
\end{abstract}


\end{frontmatter}


{\bf PROGRAM SUMMARY}

\begin{small}
\noindent
{\em Program Title:} Tomography-NMR \\
{\em CPC Library link to program files:} (to be added by Technical Editor) \\
{\em Developer's repository link:} \url{https://github.com/jiwooseo98/tomography-nmr} \\
{\em Code Ocean capsule:} (to be added by Technical Editor) \\
{\em Licensing provisions:} MIT \\
{\em Programming language:} Python 3.11+ \\
{\em Supplementary material:} Example Jupyter notebooks demonstrating complete analysis workflows for two-qubit quantum state tomography \\

{\em Nature of problem:} \\ 
Quantum state tomography is a fundamental technique for characterizing quantum systems by reconstructing their density matrices from experimental measurements.
While widely used in quantum information science, the detailed methodology for transforming raw spectroscopic data into density matrices is underdocumented in the scientific literature.
Published works typically present high-level theoretical frameworks and final results, omitting critical intermediate steps such as phase correction procedures, spectral integration methods, and systematic parameter optimization.
This lack of methodological transparency is especially problematic in nuclear magnetic resonance (NMR) quantum computing, where commercial software packages operate as closed-source systems optimized for conventional spectroscopy rather than precision quantum state reconstruction.
Researchers performing quantum information experiments require complete visibility into data processing steps, including Fourier transform implementations, peak integration algorithms, and artifact handling, in order to validate results, compare across research groups, and advance characterization methodologies.
The absence of transparent, reproducible analysis pipelines hinders progress in experimental quantum information science.
\\

{\em Solution method:} \\ 
The package provides a comprehensive open-source Python implementation for reconstructing quantum density matrices from spectroscopic data using the product operator formalism.
The analysis pipeline transforms raw time-domain signals into frequency-domain spectra via Fourier transformation with optional automatic phase correction.
Spectral peak intensities of coupled-spin doublets are extracted through three methods suited to different experimental scenarios: direct peak height measurement, numerical integration via Simpson's rule over fixed frequency windows, and systematic grid-search optimization over integration parameters (frequency offsets, integration widths, and coupling corrections) that maximizes fidelity with a known theoretical reference state.
The first two methods require no knowledge of the target state and are appropriate for reconstructing unknown quantum states; the third is a calibration and benchmarking tool for experiments where the intended state is known.
The sum and difference of left and right doublet peak intensities are mapped to product operator coefficients through a predefined tomography lookup table derived from the pulse sequence design.
From the complete set of 15 non-identity coefficients, the density matrix is reconstructed with optional enforcement of physical constraints including Hermiticity, trace normalization, and positive semi-definiteness.
State fidelity is quantified using either Fortunato's normalized projection metric or Jozsa's fidelity definition.
Uncertainty estimates are propagated through the entire pipeline.
\\

{\em Additional comments including restrictions and unusual features:} \\ 
The software is primarily designed for two-qubit systems measured via spectroscopy of coupled spin-1/2 nuclei, though the modular architecture permits extension to systems with additional spins or higher spin quantum numbers.
A prototype three-qubit operator module is included.
Input data is expected in Bruker NMR format, but the data import routines are isolated in a single preprocessing module and can be replaced for alternative spectrometer formats without modifying the reconstruction pipeline.
The package depends on standard scientific Python libraries (NumPy, SciPy, Matplotlib), the QuTiP quantum toolbox for density matrix operations, and nmrglue for Bruker format parsing.
Poetry is the recommended package manager.
The optimization routine performs systematic grid searches, which may require several minutes for comprehensive datasets but ensures reproducible, unbiased parameter selection.
Automatic phase correction is generally robust but may require manual verification for spectra with unusual lineshapes.
The software has been validated on 20 experimental quantum states with reconstruction fidelities ranging from 0.948 to 0.999.
All analysis procedures are documented in accompanying Jupyter notebooks that serve as both tutorials and templates.
\\

\end{small}


\section{Introduction}
\label{sec:introduction}

Quantum state tomography (QST) is a cornerstone experimental technique for the complete characterization of quantum systems~\cite{Nielsen2012,Chuang1997}.
By reconstructing the density matrix from a set of measurements performed in different bases, QST enables the validation of quantum gate operations, the characterization of decoherence processes, and the verification of entangled state preparation~\cite{Cory1997,Chuang1998b}.
In nuclear magnetic resonance (NMR) quantum computing, tomography has played an essential role since the earliest demonstrations of quantum algorithms on liquid-state NMR processors~\cite{Vandersypen2005,Oliveira2007}, and it remains a standard tool for benchmarking quantum operations across a variety of experimental platforms~\cite{Hradil1997, James2001, Altepeter2005}.

Despite the maturity of the theoretical framework, a significant gap exists between the formalism presented in textbooks and review articles and the practical implementation of QST on real experimental data.
Published reports typically describe the measurement protocol and present final density matrices, but omit the critical intermediate steps: how raw time-domain signals are processed into spectra, how spectral peak intensities are extracted and converted to operator coefficients, how systematic errors such as frequency offsets and phase drifts are handled, and how the reconstructed density matrix is constrained to be physical.
These details are often embedded in proprietary commercial software (e.g., Bruker TopSpin for NMR) that is optimized for conventional spectroscopy applications in chemistry and biology rather than for precision quantum state characterization.

Several open-source tools address parts of this problem individually.
The \texttt{nmrglue} library~\cite{Helmus2013} provides parsers for various spectrometer data formats, including Bruker, but does not perform any tomographic analysis.
The QuTiP framework~\cite{Lambert2026} provides general-purpose quantum object manipulation and includes generic state tomography utilities, but has no functionality for processing raw spectroscopic data or extracting operator coefficients from NMR spectra.
What is missing is an integrated, open-source pipeline that connects raw spectrometer output to validated density matrices, with every intermediate step exposed and documented.
Tomography-NMR fills this role.

The absence of transparent, well-documented analysis tools poses concrete challenges.
Researchers cannot independently verify published results without access to the same analysis tools.
Methodological improvements cannot be evaluated because the baseline procedures are undocumented.
Students and new practitioners face a steep barrier to entry, as they must reverse-engineer analysis workflows from incomplete descriptions.
Furthermore, for quantum tomography applications demanding fidelities exceeding 99\%, even subtle choices in spectral processing, such as the width of an integration window or the method of phase correction, can significantly affect the final result.

In this work, we present Tomography-NMR, an open-source Python package that addresses this gap by providing a fully transparent, reproducible implementation of quantum state tomography from spectroscopic data.
The package implements a complete pipeline from raw time-domain signals to validated density matrices, with every analysis step visible and documented.
Three integration methods are provided for different experimental scenarios: direct peak height measurement and fixed-parameter numerical integration, which require no knowledge of the target state, and a systematic optimization against a known reference state for high-precision benchmarking.
The software has been validated on 20 experimentally prepared two-qubit quantum states measured via NMR spectroscopy, achieving average reconstruction fidelities between 0.975 and 0.995.

The package is designed as a complement to our group's earlier software, PULSEE~\cite{Candoli2023}, which simulates spin dynamics and NMR observables.
While PULSEE addresses the forward problem of predicting experimental outcomes from known Hamiltonians and pulse sequences, Tomography-NMR addresses the inverse problem of reconstructing quantum states from experimental spectra.
Together, the two packages provide a complete computational framework for NMR-based quantum information experiments: PULSEE for designing and simulating pulse sequences, and Tomography-NMR for analyzing the resulting data.

The remainder of this paper is organized as follows.
Section~\ref{sec:theory} reviews the product operator formalism and establishes the connection between spectral observables and density matrix coefficients.
Section~\ref{sec:methodology} describes the experimental measurement protocol and the spectral analysis procedures.
Section~\ref{sec:software} presents the software architecture and its core modules.
Section~\ref{sec:example} provides a detailed walkthrough of a complete analysis using experimental data.
Section~\ref{sec:validation} reports validation results across 20 benchmark quantum states.
Section~\ref{sec:extensibility} discusses the extensibility of the package to other systems and platforms.
Section~\ref{sec:conclusions} summarizes the conclusions and outlines future directions.

\section{Theoretical background}
\label{sec:theory}

\subsection{Product operator formalism for two-qubit systems}
\label{sec:theory:product_ops}

We consider a system of two coupled spin-1/2 particles (qubits).
The spin angular momentum operators of qubit~1 (Q1) are denoted $\{\Iop^1_x, \Iop^1_y, \Iop^1_z\}$, and those of qubit~2 (Q2) are $\{\Iop^2_x, \Iop^2_y, \Iop^2_z\}$.
Throughout this paper, juxtaposition of a Q1 and a Q2 operator indicates a tensor product over the two-qubit Hilbert space ($\Iop^1_i \Iop^2_j \equiv \Iop^1_i \otimes \Iop^2_j$), and a single-qubit operator written alone implies a tensor product with the identity on the other qubit space ($\Iop^1_i \equiv \Iop^1_i \otimes \1$).

The two-qubit system admits 16 product operators constructed from all combinations of $\Iop^1_i$ and $\Iop^2_j$ with $i,j \in \{o, x, y, z\}$, where $\Iop_o \equiv \1$ is the identity operator.
These operators, with appropriate normalization factors, form an orthonormal basis $\Bop$ for the space of $4 \times 4$ Hermitian matrices:
\begin{equation}\label{eq:basis}
\Bop = \left\{
\1,\;
2\Iop^1_x,\; 2\Iop^1_y,\; 2\Iop^1_z,\;
2\Iop^2_x,\; 2\Iop^2_y,\; 2\Iop^2_z,\;
4\Iop^1_x\Iop^2_x,\; \ldots,\;
4\Iop^1_z\Iop^2_z
\right\},
\end{equation}
where the factors of 2 and 4 ensure orthonormality under the trace inner product:
\begin{equation}\label{eq:orthonormality}
\frac{1}{4}\text{Tr}\!\left(\Bop_{ij}\, \Bop_{kl}\right) = \delta_{ik}\,\delta_{jl}\,.
\end{equation}
Here we use the compact notation $\Bop_{ij}$ to denote the basis element proportional to $\Iop^1_i \Iop^2_j$, with the appropriate prefactor of 1, 2, or 4 understood.

\subsection{Density matrix expansion}
\label{sec:theory:dm_expansion}

Since the 16 basis operators span the entire two-qubit operator space, any density matrix $\rhoop$ can be expanded as
\begin{equation}\label{eq:rho_expansion}
\rhoop = \sum_{i,j \in \{o,x,y,z\}} c_{ij}\, \Bop_{ij}\,,
\end{equation}
where the expansion coefficients are given by
\begin{equation}\label{eq:coefficients}
c_{ij} = \frac{1}{4}\text{Tr}\!\left(\Bop_{ij}\, \rhoop\right).
\end{equation}
The normalization condition $\text{Tr}(\rhoop) = 1$, combined with the fact that the identity is the only basis element with non-zero trace, fixes $c_{oo} = 1/4$.
Consequently, the complete reconstruction of $\rhoop$ requires determining the remaining 15 non-identity coefficients $\{c_{ij}\}_{(i,j) \neq (o,o)}$ from experimental measurements.
This reduction from 16 free parameters to 15 is the starting point for the tomographic protocol described in the next section.

\subsection{Spectral observables and coefficient extraction}
\label{sec:theory:observables}

In a coupled two-spin system, each NMR spectrum exhibits a characteristic doublet structure: two peaks separated by the scalar coupling constant $J$ (in Hz).
We label the intensities of the left and right peaks as $L$ and $R$, respectively.
Following the product operator analysis of pulse-acquire experiments~\cite{Oliveira2007, Abragam2011}, the sum and difference of these peak intensities encode distinct operator coefficients.

Specifically, for the Q1 readout channel under a preparatory unitary $\hat{U}_P$, one obtains
\begin{align}
\text{Re}(L+R) &= \text{Tr}\!\left(\rhoop\;\widetilde{\Iop}^1_x\right), &
\text{Im}(L+R) &= -\text{Tr}\!\left(\rhoop\;\widetilde{\Iop}^1_y\right), \label{eq:LpR}\\
\text{Re}(L-R) &= \text{Tr}\!\left(\rhoop\; 2\widetilde{\Iop}^1_x\widetilde{\Iop}^2_z\right), &
\text{Im}(L-R) &= -\text{Tr}\!\left(\rhoop\; 2\widetilde{\Iop}^1_y\widetilde{\Iop}^2_z\right), \label{eq:LmR}
\end{align}
where the tilde denotes operators transformed by the preparatory pulse sequence, $\widetilde{\hat{A}} = \hat{U}_P^{\dagger}\, \hat{A}\, \hat{U}_P$.
Analogous expressions hold for the Q2 readout channel with the qubit labels exchanged.
Each complex spectrum therefore yields four real numbers, $\text{Re}(L+R)$, $\text{Im}(L+R)$, $\text{Re}(L-R)$, and $\text{Im}(L-R)$, each proportional to one of the 15 non-identity coefficients $c_{ij}$, determined by the choice of preparatory pulse $\hat{U}_P$.

By selecting a tomographically complete set of preparatory pulses, all 15 coefficients can be measured.
In practice, redundant measurements are obtained for many coefficients, enabling averaging to improve precision.
The specific mapping between spectra and coefficients depends on the pulse sequences used and is stored in the software as a lookup table (Section~\ref{sec:software:operators}).

\section{Methodology}
\label{sec:methodology}

\subsection{Experimental system and measurement protocol}
\label{sec:methodology:experiment}

\begin{figure}[t!]
\centering
\vspace*{-1.5cm}
\includegraphics[width=\textwidth]{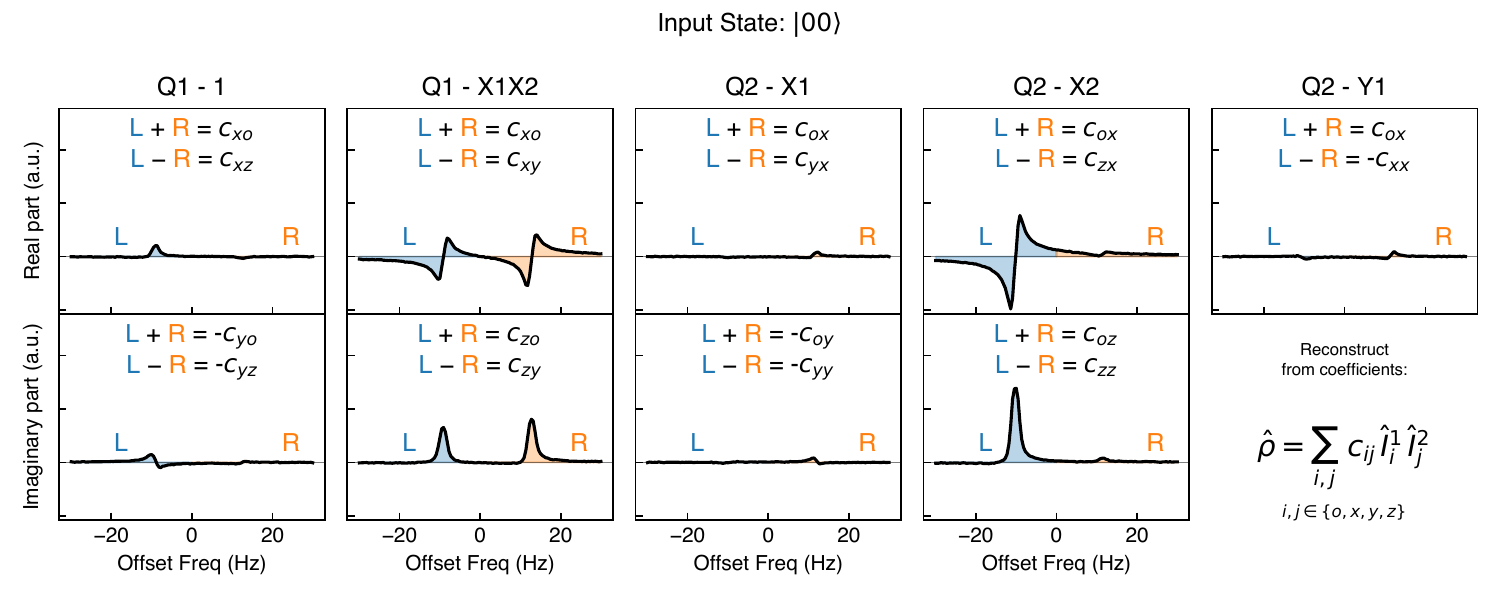}
\vspace*{-0.9cm}
\caption{Minimal set of 9 spectra sufficient to determine all 15 non-identity expansion coefficients of the two-qubit density matrix. Each spectrum displays a doublet with peaks labeled $L$ (left) and $R$ (right). The sum ($L+R$) and difference ($L-R$) of peak intensities, extracted from both the real and imaginary parts, are mapped to specific product operator coefficients as indicated. Together, these 9 spectra span the complete set of coefficients required for full state reconstruction.}
\label{fig:input_spectra}
\end{figure}

The package has been developed and validated using a two-qubit NMR system consisting of two $^{31}$P nuclei in adenosine 5'-diphosphate (ADP) dissolved in D$_2$O.
The two phosphorus spins are coupled through a scalar $J$-coupling of approximately 22~Hz.
Experiments are performed at $T = 310$~K in a magnetic field $B_0 = 14.09$~T on a Bruker NMR spectrometer.
However, the analysis methodology is not specific to this system; any coupled-spin system measured via spectroscopy can be analyzed with the same framework by adapting the data import routines.

The tomographic measurement protocol consists of seven readout pulse sequences, each applied before signal acquisition.
The readout pulses are $\pi/2$ rotations about specified axes, applied to one or both qubits, that rotate the quantum state into different measurement bases.
The seven readout configurations are: no pulse (identity), $\sqrt{X}_1$, $\sqrt{Y}_1$, $\sqrt{X}_2$, $\sqrt{Y}_2$, $\sqrt{X}_1\sqrt{X}_2$, and $\sqrt{X}_1\sqrt{Y}_2$, where $\sqrt{X}_k$ denotes a $\pi/2$ rotation of qubit $k$ about the $x$-axis.

Each readout pulse sequence produces a time-domain free induction decay (FID) signal that is Fourier-transformed into a complex frequency-domain spectrum.
The spectrum contains two doublets: one centered on the Q1 resonance frequency and one on the Q2 resonance frequency, separated by the chemical shift difference between the two $^{31}$P sites.
Each doublet is extracted separately, yielding two complex spectra per readout pulse, one for each readout channel.
Splitting each complex spectrum into its real and imaginary parts produces $7 \times 2 \times 2 = 28$ real-valued spectra in total.

\subsection{Minimal measurement set}
\label{sec:methodology:minimal}

Although the full protocol produces 28 spectra (yielding up to 56 peak intensity measurements), only 15 independent coefficients are needed to reconstruct the density matrix.
In principle, a carefully chosen subset of 9 spectra is sufficient to cover all 15 coefficients.
Figure~\ref{fig:input_spectra} illustrates one such minimal set, composed of four real--imaginary pairs from selected readout pulse and channel combinations, plus a single additional spectrum.
This reduced measurement set is useful for rapid characterization when acquisition time is limited, though the full 28-spectrum protocol provides redundant measurements that improve reconstruction accuracy through averaging.

\subsection{Spectral preprocessing}
\label{sec:methodology:preprocessing}

Raw time-domain FID signals undergo several preprocessing steps before coefficient extraction.
The Fourier transform converts each FID into a complex frequency-domain spectrum.
Digital filter artifacts near the edges of the spectral window, originating from the spectrometer's digital receiver, are identified and removed using the \texttt{nmrglue.bruker.remove\_digital\_\allowbreak{}filter()} function.
When multiple acquisitions of the same experiment are available, as is typical in pseudo-pure state preparation via temporal averaging~\cite{Cory1997, Knill1998}, the spectra are averaged to improve the signal-to-noise ratio.

Phase correction is applied to ensure that the spectral peaks are primarily absorptive (concentrated in the real channel).
The software provides both automatic and manual zeroth-order phase correction.
The automatic algorithm evaluates the integrated peak intensity at each integer phase angle in the range $[-180^\circ, 180^\circ]$ and selects the angle whose result is closest to the value predicted by the known theoretical state, with separate phase corrections determined independently for the Q1 and Q2 channels.
Because the automatic phase correction requires a known reference state, it is best suited for benchmarking experiments where the target state is specified.
For unknown states, the user may instead specify manual phase corrections based on visual inspection of the spectra.

\subsection{Peak integration methods}
\label{sec:methodology:integration}

Three methods are implemented for extracting the left ($L$) and right ($R$) peak intensities from each spectrum, listed here in order of increasing sophistication.

\subsubsection{Peak height}
\label{sec:methodology:height}

The spectral amplitude is read at the single frequency point closest to the expected peak center.
This method is computationally trivial and useful as a rapid consistency check.
However, it is sensitive to single-point noise fluctuations and discards all information about peak shape.

\subsubsection{Numerical integration with fixed parameters}
\label{sec:methodology:naive}

Simpson's rule is applied to integrate the spectrum over a frequency window of fixed width $w$ centered on each expected peak position.
The integration window width and the peak center frequencies are specified by the user based on visual inspection of the spectra.
This method provides better noise averaging than peak height and incorporates lineshape information, but its performance depends on the quality of the manually chosen parameters.

\subsubsection{Optimized integration}
\label{sec:methodology:optimized}

A systematic grid search is performed over a multi-dimensional parameter space to identify the integration settings that maximize the reconstruction fidelity.
The optimized parameters include the Q1 and Q2 frequency offsets ($\delta_1$, $\delta_2$), the integration window width ($w$), and an optional correction to the doublet splitting ($\delta_J$).
The nominal $J$-coupling value is not a free parameter but is measured directly from the observed doublet splitting in the spectra; the $\delta_J$ correction sweeps over a narrow range of $\pm 1$~Hz around this measured value to account for small calibration offsets.
For each parameter combination, the density matrix is reconstructed via Simpson integration and scored against a known theoretical reference state.
The optimization evaluates on the order of $10^3$ parameter combinations, requiring typically 2--5 minutes on modern hardware.
Because every combination of swept parameters is evaluated, this cost grows quickly if the grid resolution is increased or if additional parameters must be swept, as would be the case for systems with more coupled spins.
Extending the optimized method to larger systems would therefore likely require a more efficient search strategy in place of the exhaustive grid search (see Section~\ref{sec:extensibility}).
Results are cached for subsequent analysis runs.

This method requires knowledge of the target state and is therefore a benchmarking and calibration tool, not a general-purpose reconstruction method.
For reconstructing unknown or ``mystery'' states where the theoretical target is not available, the fixed-parameter integration method (Section~\ref{sec:methodology:naive}) is the appropriate choice: the user sets integration parameters by visual inspection of the spectra and obtains a reconstruction without any reference to theory.
The peak height method (Section~\ref{sec:methodology:height}) serves as a rapid consistency check.
Together, the three methods form a hierarchy suited to different experimental scenarios, from quick diagnostics to high-precision benchmarking of known states.

\subsection{Density matrix reconstruction and physical constraints}
\label{sec:methodology:reconstruction}

From the complete set of operator coefficients $\{c_{ij}\}$ obtained across all pulse sequences and readout channels, the density matrix is assembled via Eq.~\eqref{eq:rho_expansion}.
When multiple spectra yield the same coefficient (due to measurement redundancy), the values are averaged and their standard deviation provides an uncertainty estimate.

The raw reconstructed density matrix is traceless, since all non-identity basis operators have zero trace, and generally possesses negative eigenvalues, making it not yet a physically valid quantum state.
This is expected: NMR experiments measure the deviation density matrix (the traceless part of $\rhoop$), because the signal is proportional to the magnetization relative to thermal equilibrium, not to the full state.
The identity component, which carries no observable NMR signal, must therefore be restored by construction.
A cleanup procedure enforces the required physical constraints:
\begin{enumerate}
\item \textbf{Trace normalization:} An identity component $\frac{1}{4}\1$ is added so that $\text{Tr}(\rhoop) = 1$, restoring the full density matrix from the measured deviation matrix.
\item \textbf{Hermiticity:} Satisfied by construction, since the basis operators are Hermitian and the coefficients are real.
\item \textbf{Positive semi-definiteness:} Optionally enforced by truncating negative eigenvalues via eigendecomposition and renormalizing. Small negative eigenvalues arise from experimental noise in the measured coefficients and do not indicate a failure of the reconstruction procedure. This is the recommended cleanup mode and typically yields the highest reconstruction fidelity.
\end{enumerate}

\subsection{Fidelity metrics}
\label{sec:methodology:fidelity}

The quality of a reconstructed state $\rhom$ is quantified by its fidelity with the corresponding theoretical target $\rhoth$.
Two metrics are implemented.

Fortunato's projection~\cite{Fortunato2002} provides a normalized trace inner product:
\begin{equation}\label{eq:fortunato}
F_{\text{proj}}(\rhom, \rhoth) = \frac{\text{Tr}(\rhom\, \rhoth)}{\sqrt{\text{Tr}(\rhom^2)\;\text{Tr}(\rhoth^2)}}\,.
\end{equation}
For pure target states ($\rhoth^2 = \rhoth$), this reduces to $\text{Tr}(\rhom\, \rhoth) / \sqrt{\text{Tr}(\rhom^2)}$, which provides good numerical stability during optimization.

Jozsa's fidelity~\cite{Jozsa1994} is defined as
\begin{equation}\label{eq:jozsa}
F_J(\rhom, \rhoth) = \left[\text{Tr}\sqrt{\sqrt{\rhom}\;\rhoth\;\sqrt{\rhom}}\right]^2,
\end{equation}
which is the standard definition used widely in quantum information theory.
For nearly pure states, the two metrics converge and yield consistent results, serving as a useful cross-check.
Both are implemented in the software, and the user may select either as the optimization objective.

\section{Software architecture}
\label{sec:software}

Tomography-NMR is written entirely in Python~3.11+ and uses Poetry for dependency management.
The package depends on NumPy for array operations and linear algebra, SciPy for numerical integration, Matplotlib for visualization, QuTiP~\cite{Lambert2026} for quantum object manipulation, and nmrglue~\cite{Helmus2013} for reading Bruker NMR data formats.
The source code is organized into modular components that separate data import, spectral analysis, density matrix reconstruction, and visualization.

\subsection{\texttt{operators.py}: Operator basis and quantum states}
\label{sec:software:operators}

This module defines the complete product operator basis for a two-qubit system as QuTiP \texttt{Qobj} objects in the $4 \times 4$ Hilbert space.
Single-qubit spin operators are labeled \texttt{Ix}, \texttt{Iy}, \texttt{Iz} for Q1 and \texttt{Sx}, \texttt{Sy}, \texttt{Sz} for Q2, with two-qubit products written as \texttt{IxSy}, \texttt{IzSx}, etc.
The module also provides predefined density matrices for all four computational basis states and four Bell states, both in their physical form ($\text{Tr}(\rhoop) = 1$) and as traceless deviation matrices.

The central data structure is the tomography lookup table \texttt{product\_\allowbreak{}operators}: a length-28 list that maps each expanded spectrum index to its corresponding pair of product operators (the $(L+R)$ and $(L-R)$ operators).
This table encodes the connection between spectral measurements and density matrix coefficients established in Eqs.~\eqref{eq:LpR}--\eqref{eq:LmR}, and its correctness is verified by the utility function \texttt{check\_basis\_complete()}, which confirms that all 15 non-identity operators are covered.

Additionally, the module defines standard quantum gates (single-qubit rotations \texttt{Rx}, \texttt{Ry}, \texttt{Rz}, the Hadamard gate, the CNOT gate, and the $T$ phase gate) as well as the $J$-coupling evolution operator, enabling the construction of arbitrary pulse sequences for theoretical reference state calculations.

\subsection{\texttt{preprocessing.py}: Data import and spectral processing}
\label{sec:software:preprocessing}

This module reads raw acquisition data from Bruker NMR format files and produces frequency-domain complex spectra ready for tomography.
The function \texttt{process\_data()} walks numbered acquisition subfolders, Fourier-transforms each time-domain signal, optionally removes digital filter artifacts, temporally averages repeated acquisitions, and applies automatic or manual phase correction.
It returns a frequency axis (in Hz) and a complex array of shape $(n_\text{readouts}, n_\text{points})$, typically $(14, 3130)$ for a full two-qubit tomography set.

The downstream analysis pipeline is not specific to NMR; only this preprocessing module is tied to the Bruker data layout.
Adapting the package to a different spectrometer or measurement modality requires only replacing the data-loading step within this module.

\subsection{\texttt{integrate.py}: Peak integration and optimization}
\label{sec:software:integrate}

This module provides the three integration methods described in Section~\ref{sec:methodology:integration}.
The function \texttt{integrate\_spectra()} implements fixed-parameter Simpson integration: it integrates frequency windows around each doublet peak, assembles the $(L+R)$ and $(L-R)$ coefficients via the tomography lookup table, and reconstructs the density matrix through the \texttt{Coefficient\allowbreak{}Groups} class.
Optional parameters control frequency offsets, spectrum exclusion, cleanup method, and error propagation.

The function \texttt{integrate\_optimized()} performs the grid-search optimization described in Section~\ref{sec:methodology:optimized}.
For each parameter combination, it calls \texttt{integrate\_spectra()} and scores the result against a theoretical reference using the selected fidelity metric.
It returns the best density matrix, fidelity, and parameter set, with optional uncertainty estimates.

\subsection{\texttt{coeff\_groups\_class.py}: Coefficient management and reconstruction}
\label{sec:software:coeff_groups}

The \texttt{CoefficientGroups} dataclass collects, organizes, and averages the product operator coefficients extracted from spectral measurements.
Coefficients are added one at a time via \texttt{add\_coefficient()}, which records the operator, the measured value, and the readout channel label.
The method \texttt{reconstruct\_rho()} averages redundant measurements for each operator, assembles the density matrix from the weighted sum of basis operators, and applies the cleanup procedure via \texttt{clean\_dm()}.
The method \texttt{get\_error()} propagates the standard deviation of redundant coefficient measurements into an element-wise error matrix for the reconstructed density matrix.
This captures the statistical scatter among repeated measurements of the same coefficient but does not account for systematic effects, such as slow phase drift over the course of an experiment, that would shift redundant measurements consistently in the same direction.

\subsection{\texttt{fidelity.py}: State comparison metrics}
\label{sec:software:fidelity}

This module implements both the Fortunato projection (Eq.~\eqref{eq:fortunato}) and the Jozsa fidelity (Eq.~\eqref{eq:jozsa}), along with error propagation utilities for computing uncertainty in the fidelity value.
A convenience function \texttt{is\_positive\_\allowbreak{}semidefinite()} checks whether a density matrix has all non-negative eigenvalues.

\subsection{Additional modules}
\label{sec:software:additional}

The \texttt{phase.py} module implements automatic zeroth-order phase correction by optimizing the phase angle to maximize the real integrated signal.
The \texttt{uncertainty.py} module provides error propagation utilities for matrix products and trace operations, used internally by the fidelity and reconstruction routines.
The \texttt{visualize.py} module generates publication-quality density matrix bar plots (with bar height encoding magnitude and color encoding complex phase), spectral overlay plots with integration window annotations, and various diagnostic visualizations.
The \texttt{global\_constants.py} module centralizes project-wide definitions including file paths, display settings, color codes, and the ordered list of spectrum labels.

A prototype three-qubit extension is provided in \texttt{operators\_3spin.py}, which defines the complete set of 63 non-identity product operators for three spin-1/2 particles, along with single-qubit rotation methods and a three-spin coupling unitary.
This module is not used in the current two-qubit pipeline but serves as a starting point for extensions to larger systems.

\begin{figure}[t]
\vspace*{-0.5cm}
\centering
\includegraphics[width=0.45\textwidth]{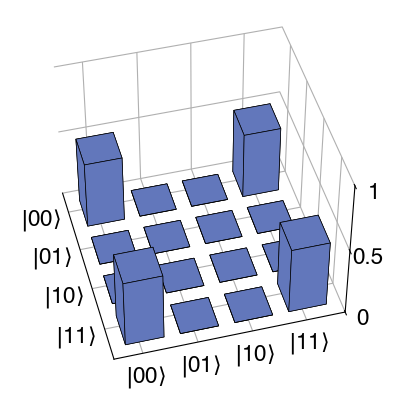}
\vspace*{-0.30cm}
\caption{Theoretical density matrix for the $\ket{00}$ computational basis state, visualized as a 3D bar chart. The single non-zero diagonal element at $\ket{00}\!\bra{00}$ confirms the pure state $\rhoop = \ket{00}\!\bra{00}$.}
\label{fig:theory_rho}
\end{figure}

\section{Example analysis}
\label{sec:example}

We demonstrate the complete analysis workflow using experimental data for the $\ket{00}$ computational basis state, prepared via temporal averaging of pseudo-pure state preparation sequences~\cite{Cory1997,Knill1998}.
The accompanying Jupyter notebook (\texttt{analysis\_example.ipynb}) provides a step-by-step template that users can adapt to their own datasets by changing the data folder path and state name.

\subsection{Configuration and theoretical reference}
\label{sec:example:config}

The analysis begins by importing the relevant modules and setting configuration flags that control various aspects of the reconstruction:
\begin{lstlisting}
state_name = "1000"         # Basis state |00>
AUTO_PHASE = True           # Automatic phase correction
TRUNC_NEG = True            # Enforce positive semi-definiteness
COMPARE_RAW = False         # Use cleaned density matrix
COMPARE_SIMULATION = False  # Compare to ideal theory
\end{lstlisting}
\textbf{Notes on naming convention:} The four-character state names \texttt{"1000"}, \texttt{"0100"}, \texttt{"0010"}, and \texttt{"0001"} correspond to the two-qubit computational basis states $\ket{00}$, $\ket{01}$, $\ket{10}$, and $\ket{11}$, respectively.
Each name represents the diagonal elements of the density matrix (e.g., \texttt{"1000"} denotes the state with diagonal $\{1,0,0,0\}$ of the density matrix $\ket{00}\!\bra{00}$).
This convention was chosen to make the correspondence to the density matrix explicit.
Some newer variables and files in the codebase use the shorter two-character labels \texttt{"00"}, \texttt{"01"}, \texttt{"10"}, and \texttt{"11"} instead, corresponding to the ``ket'' naming convention.
Both the ``diagonal'' and ``ket'' namings are used but refer to the same states; readers adapting the accompanying notebooks should be aware that both appear in the codebase.
The theoretical target state is loaded from the predefined operator definitions:
\begin{lstlisting}
rho_theory = op.CLEAN_1000  # |00><00|, trace = 1
\end{lstlisting}

The theoretical density matrix is visualized in Fig.~\ref{fig:theory_rho} for reference.

\begin{figure}[h!]
\vspace*{-0.1cm}
\centering
\includegraphics[width=0.80\textwidth]{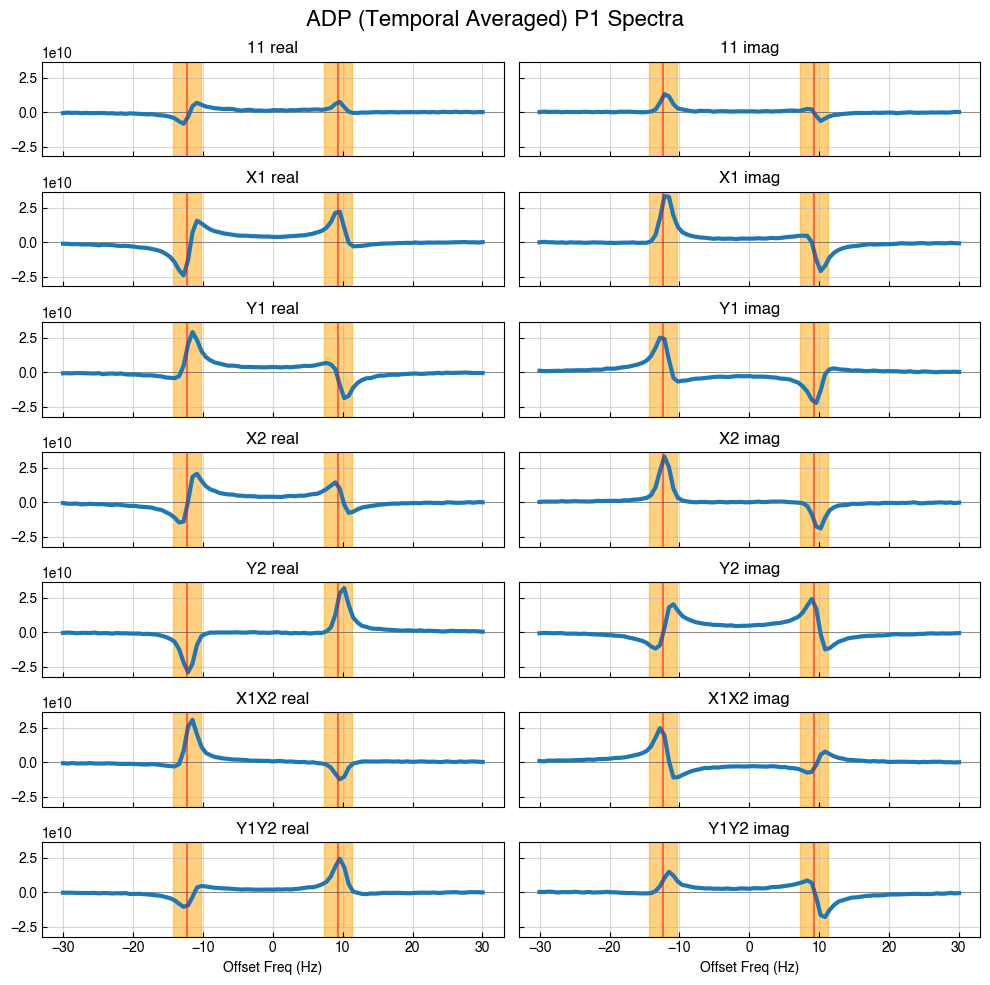}
\vspace*{-0.3cm}
\caption{Q1 readout channel spectra for all 14 tomographic measurements (7 readout pulses $\times$ 2 real/imaginary parts). Red vertical lines mark integration centers; orange shaded regions indicate integration windows of width $w = 4$~Hz. Each spectrum displays the characteristic doublet separated by $J \approx 22$~Hz.}
\label{fig:q1_spectra}
\end{figure}

\subsection{Data loading and preprocessing}
\label{sec:example:loading}

Raw time-domain NMR data is loaded from Bruker-format acquisition folders and processed into frequency-domain spectra:
\begin{lstlisting}
freqs, all_spectra_complex = preprocessing.process_data(
    data_folder, remove_digital=True,
    auto_phase=AUTO_PHASE, state_name=state_name)
\end{lstlisting}
This function performs Fourier transformation, removes digital filter artifacts, temporally averages multiple acquisitions, and applies automatic phase correction.
The resulting complex spectra are split into real and imaginary parts, yielding 28 real-valued spectra (14 per readout channel).

\begin{figure}[h!]
\vspace*{-0.1cm}
\centering
\includegraphics[width=0.8\textwidth]{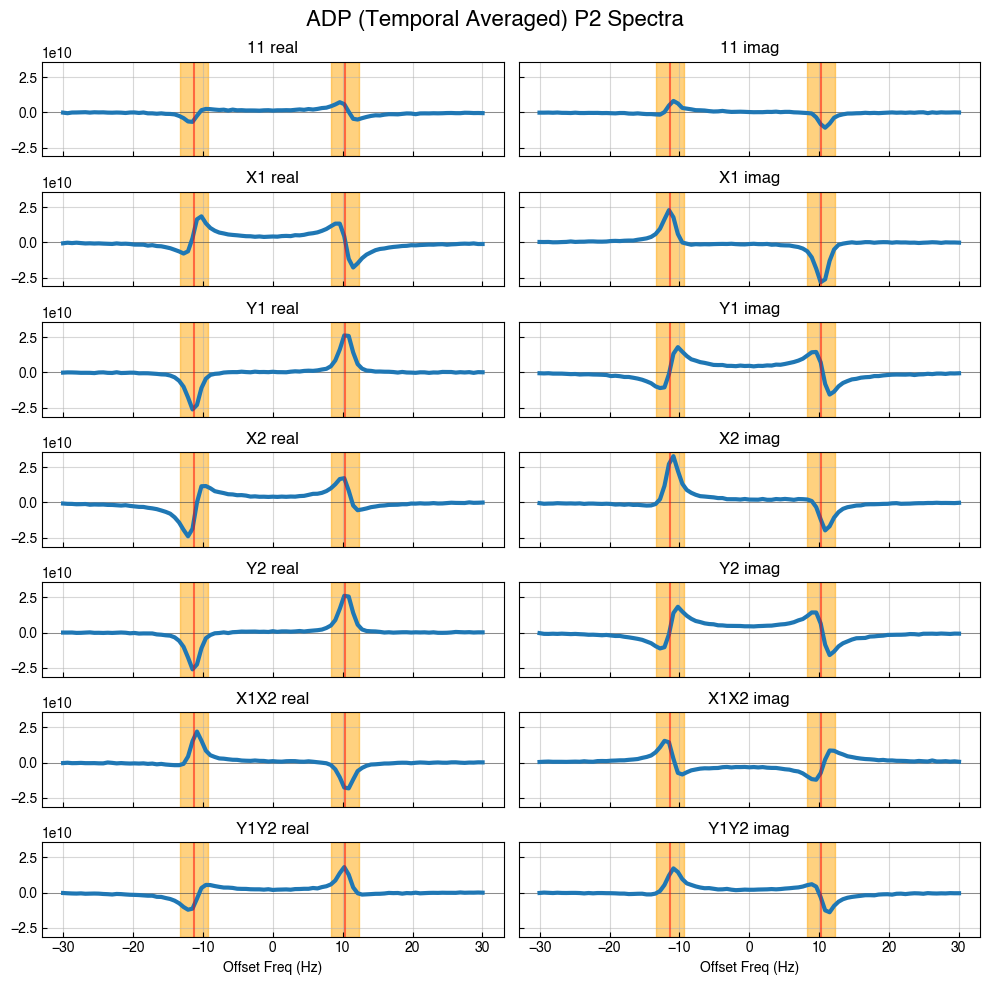}
\vspace*{-0.3cm}
\caption{Q2 readout channel spectra for all 14 tomographic measurements, with independently adjusted frequency offset. The Q2 channel requires a different offset than Q1 due to slight chemical shift differences between the two $^{31}$P nuclei.}
\label{fig:q2_spectra}
\end{figure}

The $J$-coupling constant is measured from the observed doublet splitting:
\begin{lstlisting}
J = 22      # Hz, from observed peak separation
L_FREQ = -J / 2
R_FREQ =  J / 2
\end{lstlisting}

Before integration, all spectra are visualized with overlay markers indicating the integration centers (vertical lines) and integration windows (shaded regions).
Figure~\ref{fig:q1_spectra} shows the 14 Q1 spectra, each displaying the characteristic doublet structure.
The Q1 and Q2 frequency offsets (\texttt{p1\_offset}, \texttt{p2\_offset}) and integration window width (\texttt{INT\_WIDTH}) are adjusted manually based on visual inspection to ensure that integration windows capture the peaks accurately:
\begin{lstlisting}
INT_WIDTH = 4       # Integration window width (Hz)
p1_offset = 1.5     # Q1 frequency offset (Hz)
p2_offset = 0.5     # Q2 frequency offset (Hz)
\end{lstlisting}

Figure~\ref{fig:q2_spectra} shows the corresponding Q2 spectra with independently adjusted offset parameters.
These visualizations enable quality assessment of spectral phasing, baseline flatness, and peak alignment before proceeding to integration.

\subsection{Spectral visualization and parameter adjustment}
\label{sec:example:visualization}

\subsection{Density matrix reconstruction}
\label{sec:example:reconstruction}

The three integration methods described in Section~\ref{sec:methodology:integration} are applied sequentially.
For the peak height method, spectral amplitudes are read at the expected peak frequencies and assembled into operator coefficients:
\begin{lstlisting}
coeff_groups_height = CoefficientGroups()
for (i, spectrum) in enumerate(all_spectra):
    loc_L, loc_R = p1_freqs if i < 14 else p2_freqs
    L = spectrum[np.absolute(freqs - loc_L).argmin()]
    R = spectrum[np.absolute(freqs - loc_R).argmin()]
    coeff_groups_height.add_coefficient(
        op.product_operators[i][0], L + R,
        integrate.index_to_element(i))
    coeff_groups_height.add_coefficient(
        op.product_operators[i][1], L - R,
        integrate.index_to_element(i))
rho_height = coeff_groups_height.reconstruct_rho(
    trunc_neg=TRUNC_NEG, add_identity=True)
\end{lstlisting}

\begin{figure}[t!]
\centering
\vspace*{-1.5cm}
\includegraphics[width=0.99\textwidth]{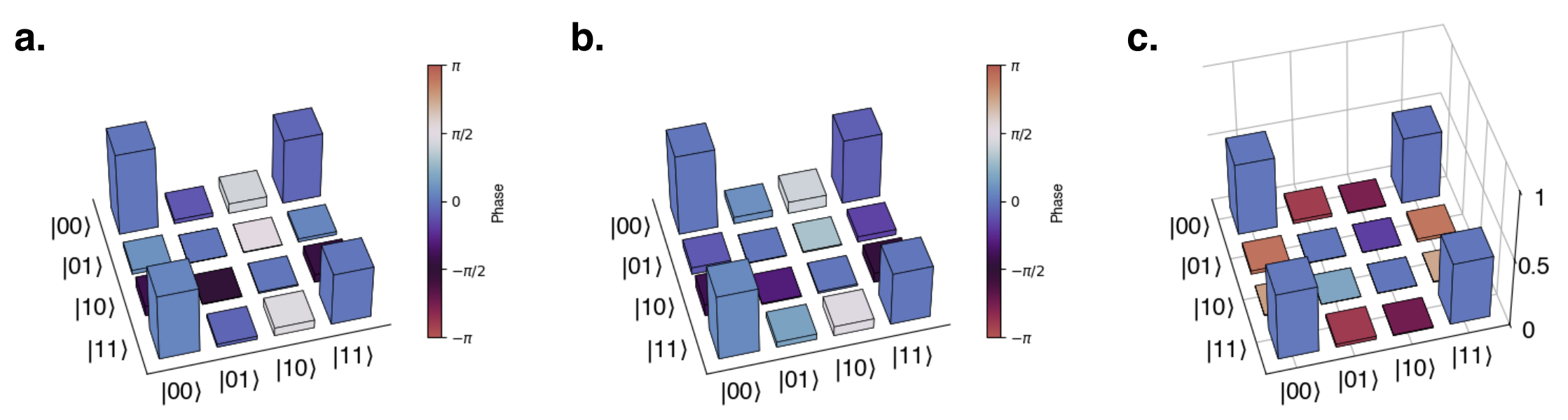} 
\vspace*{-0.2cm}
\caption{Comparison of the density matrix reconstruction from different integration methods. {\bf a.} Density matrix reconstructed using the peak height method ($F = 0.980$). While the dominant $\ket{00}\!\bra{00}$ element is correctly identified, minor deviations are visible in off-diagonal and subdominant diagonal elements.
{\bf b.} Density matrix from fixed-parameter Simpson integration ($F = 0.980$). The result is similar to the peak height method, with small differences arising from the noise-averaging effect of finite integration windows.
{\bf c.} Density matrix from optimized integration ($F = 0.998$). The systematic parameter optimization significantly improves reconstruction accuracy, yielding a density matrix that closely matches the theoretical $\ket{00}$ state. Residual off-diagonal elements of order $10^{-2}$ are consistent with experimental imperfections.}
\label{fig:height_rho}
\end{figure}

For the fixed-parameter integration, a single function call performs Simpson integration over all spectra:
\begin{lstlisting}
rho_naive = integrate.integrate_spectra(
    freqs, all_spectra, p1_freqs, p2_freqs, INT_WIDTH,
    trunc_neg=TRUNC_NEG, add_identity=True)
\end{lstlisting}

For the optimized integration, a grid search over parameters maximizes the reconstruction fidelity:
\begin{lstlisting}
rho_opt, best_fidelity, best_params, rho_error = \
    integrate.integrate_optimized(
        freqs, all_spectra, p1_freqs, p2_freqs,
        rho_theory, return_error=True,
        j_offset_range=np.linspace(-1, 1, 21),
        trunc_neg=TRUNC_NEG)
\end{lstlisting}



The results of all three methods are summarized in Table~\ref{tab:methods_comparison} and visualized in Fig.~\ref{fig:height_rho}. 
The progression from peak height through fixed integration to optimized integration demonstrates the critical importance of systematic parameter optimization for achieving high-fidelity reconstruction.

\begin{table}[h!]
\centering
\caption{Comparison of the three integration methods for the $\ket{00}$ basis state reconstruction.}
\label{tab:methods_comparison}
\begin{tabular}{lcc}
\toprule
Method & Fidelity & Computation time \\
\midrule
Peak height         & 0.980 & $<$ 1 s \\
Fixed integration   & 0.980 & $<$ 1 s \\
Optimized integration & \textbf{0.998} & $\sim$2 min \\
\bottomrule
\end{tabular}
\end{table}

\subsection{Comparison with simulated states}
\label{sec:example:simulation}

The ideal theoretical states assume instantaneous pulses with perfect rotation angles.
In practice, $J$-coupling evolution occurs during finite-duration pulses, and residual calibration errors produce imperfect rotations.
The package includes a unitary simulation notebook (\texttt{unitary\_simulation\_with\_J.\allowbreak{}ipynb}) that accounts for these non-idealities.
By comparing the experimentally reconstructed state against both the ideal and simulated theoretical states, one can distinguish genuine experimental noise from systematic deviations predicted by the model.
This comparison is enabled by setting the \texttt{COMPARE\_SIMULATION} flag, which loads a pre-computed simulated density matrix as the reference for optimization.

\section{Validation}
\label{sec:validation}

The package has been validated on 20 experimentally prepared two-qubit quantum states measured on the ADP system described in Section~\ref{sec:methodology:experiment}.
The states span five categories of increasing complexity: four computational basis states ($\ket{00}$, $\ket{01}$, $\ket{10}$, $\ket{11}$), four Hadamard gate outputs, four CNOT gate outputs, four Bell states (prepared by Hadamard followed by CNOT), and four Hadamard + $T$ phase gate outputs.
All reconstructions use the optimized integration method with positive semi-definiteness enforcement.

Table~\ref{tab:validation} summarizes the average Fortunato projection fidelity for each state category.
Individual state fidelities range from 0.948 to 0.999.

\begin{table}[h!]
\centering
\caption{Validation results for 20 experimentally prepared two-qubit states. The Fortunato projection fidelity is averaged over all states within each category.}
\label{tab:validation}
\begin{tabular}{lcc}
\toprule
State category & States tested & Average fidelity \\
\midrule
Computational basis states    & 4 & 0.995 \\
Hadamard gate outputs         & 4 & 0.992 \\
CNOT gate outputs             & 4 & 0.983 \\
Bell states (H + CNOT)        & 4 & 0.980 \\
H + $T$ gate outputs          & 4 & 0.975 \\
\bottomrule
\end{tabular}
\end{table}

Several trends are evident.
The computational basis states, which are diagonal in the measurement basis, achieve the highest fidelities.
States involving multi-qubit gates (CNOT, H+CNOT) show slightly lower fidelities, consistent with the cumulative effect of pulse imperfections in longer gate sequences.
The H+$T$ gate outputs exhibit the lowest average fidelity, likely reflecting the additional complexity of implementing the $T$ phase gate.
Across all categories, the average fidelity exceeds 0.975, demonstrating that the software reliably reconstructs quantum states with high accuracy.

These fidelities are obtained using the optimized integration method, which tunes integration parameters against the known theoretical state for each preparation (Section~\ref{sec:methodology:optimized}).
They therefore represent self-consistent benchmarks: they confirm that the analysis pipeline can recover the expected state when the spectrometer parameters are optimally calibrated, but they do not constitute blind reconstructions.
For comparison, the fixed-parameter integration method, which uses manually chosen integration settings and does not reference the theoretical state, achieves fidelities of approximately 0.980 for the same data (see Table~\ref{tab:methods_comparison}).
This provides an independent, theory-free lower bound on reconstruction quality and is the appropriate method for characterizing unknown quantum states.

Uncertainty estimates, propagated through the analysis pipeline via the \texttt{uncertainty.py} module, are typically of order $10^{-3}$ to $10^{-4}$ for individual density matrix elements.
The uncertainty in the fidelity value, computed by the \texttt{fortunato\_error()} function, is correspondingly small ($\sim 10^{-3}$).

\section{Extensibility}
\label{sec:extensibility}

Although developed and validated on a two-qubit NMR system, the package is designed with extensibility in mind.
The underlying framework (product operator expansion, coefficient extraction from spectral peak intensities, and density matrix reconstruction with physical constraints) is general and applies to any spin system where spectral data encode information about the quantum state.

\textbf{Higher-dimensional systems.}
Extension to three or more qubits requires constructing a larger operator basis ($4^N - 1$ non-identity operators for $N$ qubits) and a correspondingly larger tomography lookup table.
The prototype module \texttt{operators\_3spin.py} already provides the complete set of 63 non-identity product operators for three spin-1/2 particles, along with single-qubit rotations and a pairwise coupling evolution operator parameterized by three coupling constants ($J_{12}$, $J_{13}$, $J_{23}$).
The number of required measurements scales exponentially with system size, which is a fundamental limitation of full state tomography.
Beyond measurement count, spectral crowding presents a practical challenge: as the number of coupled spins increases, multiplet peaks begin to overlap, and the simple doublet integration used in the current two-qubit pipeline would need to be replaced by spectral deconvolution methods.
The current implementation does not include such deconvolution, and extending the software to three or more qubits remains an open problem requiring both algorithmic development and careful experimental validation.
The present package is therefore intentionally scoped to the two-qubit regime, where NMR-based tomography remains experimentally viable at high fidelity, rather than offering a general $N$-qubit solution.

\textbf{Higher spin quantum numbers.}
For nuclear or electronic spins with $S > 1/2$, the operator basis generalizes to $(2S+1)^2 - 1$ traceless generators per spin.
QuTiP's \texttt{jmat(S)} functions provide the necessary spin-$S$ operators, enabling straightforward construction of the appropriate basis.

\textbf{Alternative spectroscopic platforms.}
The same tomographic approach applies to electron spin systems measured via pulsed electron paramagnetic resonance (EPR), where the spectral readout is analogous to NMR.
More broadly, any spectroscopic measurement that produces peaks whose intensities encode operator expectation values can be analyzed within this framework.

\textbf{Other spectrometer formats.}
Data loading is isolated in \texttt{preproce-\allowbreak{}ssing.py}.
Replacing the Bruker reader with a parser for Varian, Agilent, JEOL, or any other spectrometer format requires no modifications to the reconstruction pipeline.

\textbf{Advanced optimization.}
The current grid-search optimization can be replaced with gradient-based methods, Bayesian optimization, or evolutionary algorithms for improved efficiency.
Such methods would also be necessary for scaling to larger systems, where the grid search becomes computationally prohibitive due to the increased dimensionality of the parameter space.
The modular design of \texttt{integrate.py} allows custom optimization strategies to be incorporated without restructuring the codebase.

\section{Conclusions}
\label{sec:conclusions}

We have presented Tomography-NMR, an open-source Python package for quantum state tomography from spectroscopic measurement data.
The package addresses a practical gap in the quantum information literature by providing a fully transparent, documented implementation of the complete analysis pipeline, from raw time-domain signals to validated density matrices with quantified uncertainties.

The software achieves reconstruction fidelities exceeding 99\% for benchmark quantum states when integration parameters are optimized against known target states, and fidelities of approximately 98\% using theory independent fixed-parameter integration.
Validation on 20 experimentally prepared two-qubit states demonstrates consistent high-accuracy reconstruction across computational basis states, entangled Bell states, and various quantum gate outputs.

By making the analysis methodology transparent and reproducible, this package enables independent verification of published results, facilitates comparison across research groups, and provides new practitioners with a well-documented entry point into experimental quantum state tomography.
The modular architecture is designed to support adaptation to alternative spectroscopic platforms and higher spin numbers, though extension to systems with more than two qubits will require additional algorithmic work to handle spectral crowding and optimization scalability.

Together with the PULSEE simulation software~\cite{Candoli2023}, this package provides a complete open-source computational framework for NMR-based quantum information experiments, spanning both the forward problem of simulating experimental outcomes and the inverse problem of reconstructing quantum states from data.

Future development directions include extending support to three-qubit and higher-dimensional systems, implementing real-time analysis capabilities for adaptive experimental protocols, incorporating Bayesian optimization to enable more efficient parameter searches, and developing additional validation metrics, such as entanglement witnesses and extensions of quantum process tomography.

\section*{CRediT authorship contribution statement}

\textbf{Jiwoo Seo:} Software, Investigation, Validation, Data curation, Visualization, Writing -- original draft.
\textbf{Vesna F. Mitrovi\'{c}:} Conceptualization, Methodology, Supervision, Resources, Writing -- review \& editing.

\section*{Declaration of competing interest}

The authors declare that they have no known competing financial interests or personal relationships that could have appeared to influence the work reported in this paper.

\section*{Data availability}

The software source code and example analysis notebooks are available at \url{https://github.com/jiwooseo98/tomography-nmr}.
Upon acceptance, the source code will be archived on Zenodo with an assigned DOI to ensure long-term availability.
Experimental datasets used for validation are available upon request.

\section*{Acknowledgements}

We thank Ilija K. Nikolov, Adrien Rosuel, and Donovan Davino  for their collaboration and valuable insights during the development of this package.
V.F.M.\ acknowledges support from Brown University Seed Funds.

\bibliographystyle{elsarticle-num}
\bibliography{tomography.bib}

\end{document}